\documentclass
[preprint,showpacs,10pt,nobibnotes,noshowkeys,unsortedaddress,pra,twocolumn,nofootinbib,notitlepage,twoside]{revtex4}%
\usepackage{amsfonts}
\usepackage{amsmath}
\usepackage{amssymb}
\usepackage{graphicx}%
\begin{document}
\title{Cluster-state preparation and multipartite entanglement analyzer with fermions}
\author{X. L. Zhang$^{1,2}$}
\email{xili-zhang@hotmail.com}
\author{M. Feng$^{1}$}
\author{K. L. Gao$^{1}$}
\affiliation{$^{1}$State Key Laboratory of Magnetic Resonance and Atomic and Molecular
Physics, Wuhan Institute of Physics and Mathematics, Chinese Academy of
Sciences, Wuhan 430071, People's Republic of China}
\affiliation{$^{2}$Graduate School of the Chinese Academy of Science, Bejing 100049,
People's Republic of China}


\begin{abstract}
Quantum cluster states and entangled state analyzers are essential to
measurement-based quantum computing. We propose to generate a quantum
cluster-state and to make multipartite entanglement analyzer by using
noninteracting free electrons or conduction electrons in quantum dots, based
on polarizing beam splitters, charge detectors and single-spin rotations. Our
schemes are deterministic without the need of qubit-qubit interaction.

\end{abstract}
\maketitle

Quantum computers own great advantages over existing computers to solve
classically intractable problems and to speed up some tractable solutions
\cite{1,2,3}. Most current researches are exploring quantum computing based on
externally controlled qubit-qubit interaction. While there are a kind of
alternative quantum computing proposals based on qubit measurements.

The quantum cluster state \cite{4}, a special multipartite entangled state, is
the key ingredient in a measurement-based quantum computing, i.e., one-way
quantum computing \cite{5}. With a cluster state of an array of qubits, we can
carry out expected quantum gates by some single-qubit operations and
detections. In a lattice structure, for any site $a$ of the lattice, the
cluster state $\left\vert \phi_{\left\{  k\right\}  }\right\rangle _{C}$ obeys
the set of eigenvalue equations $K^{\left(  a\right)  }\left\vert
\phi_{\left\{  k\right\}  }\right\rangle _{C}=(-1)^{\kappa_{a}}\left\vert
\phi_{\left\{  k\right\}  }\right\rangle _{C},$ with the corresponding
operator $K^{\left(  a\right)  }=\sigma_{x}^{\left(  a\right)  }\underset{b\in
nghb\left(  a\right)  }{\otimes}\sigma_{z}^{\left(  b\right)  },$ where
nghb(a) specifies the sites of all the neighbors of the site $a$ and
$\kappa_{a}\in\left\{  0,1\right\}  $. The operators $K^{\left(  a\right)  }$
with \{$a\in$ lattice\} form a complete family of commuting operators on the
lattice. It has been shown that some physical systems such as quantum dots,
optical photons and cavity quantum electrodynamics (QED) are suitable for
preparation of cluster states \cite{6,7,8,9}.

Alternatively, if we have Bell - state measurements (or say, analyzers) and
some initial sources of entanglements, quantum gates necessary for universal
quantum computing are also available. A typical example for this quantum
computing is to use linear optical elements, e.g., polarizing beam splitters,
as proposed in \cite{10,11}. It is generally believed that for
measurement-based quantum computing, either full Bell-state measurements in
combination with some initial source of entanglement or partial Bell-state
measurements are sufficient for a universal quantum computing \cite{12,13}.

In this work, we report how to generate cluster states and to make analyzers
for multipartite entangled states with noninteracting free electrons or
conduction electrons in quantum dots, by using beam splitters and single-spin
rotations. Our work is inspired by recent work \cite{12,14}, which show that
quantum information processing with such electrons could be simpler but more
powerful than by using photons. However, a previous no-go theorem \cite{15,16}
declared that the operators with beam splitters and single-spin rotations, can
only achieve universal quantum computing in noninteracting boson systems
instead of in fermions'. Nevertheless, Ref. \cite{17} shows that entangled
states can be created probabilistically using beam splitters and position
detections for bosons as well as for fermions. Recently, more specific schemes
\cite{12,14} were proposed by using noninteracting free electrons or
conduction electrons in quantum dots to carry out universal quantum computing
with beam splitters and single-spin rotations. The present paper consists of
two parts. In the first part, we first describe how to construct the cluster
states with noninteracting electrons using polarizing beam splitters and
single-spin rotations, assisted by parity check on the qubits. In contrast to
the design in \cite{18} with success probability of 0.25 by using polarized
photons, our scheme is deterministic without using any prior entanglement
necessary in former parametric-down-conversion-based protocols. In the second
part of the paper, we describe entanglement analyzers for multipartite states
which would be useful for quantum communication with multipartite states and
for judging multipartite entanglement \cite{19,20}\textit{. }

Like in [12,14], we encode the qubits into the spin degrees of freedom of the
noninteracting electrons. We first show in Fig. 1 how to realize a controlled
phase (C-phase) gate using the encoder designed in \cite{12}, which is
modified from the Fig. 3 of \cite{12}. Based on this design, we could realize
a cluster state in the following. We consider N noninteracting electrons in
Fig. 2 prepared initially in a product state $\left\vert \phi_{0}\right\rangle
=\otimes_{j}\left\vert +\right\rangle _{j}$, where indexes $j=1,2,3,\cdots$
refer to the sites of the atoms and $\left\vert +\right\rangle _{j}$ is the
eigenstate of $\sigma_{x}^{(j)}$ with the eigenvalue 1. A straightforward
deduction yields that, when $P_{j}=1,$ the pair states of 1-2, 3-4, and so on,
will collapse to
\begin{equation}
\frac{1}{\sqrt{2}}\left(  \left\vert 00\right\rangle +\left\vert
11\right\rangle \right)  .
\end{equation}
After Hadamard transformations are performed on the second qubits of each
pairs, Eq. (1) becomes%
\begin{equation}
\frac{1}{2}\left(  \left\vert 00\right\rangle +\left\vert 01\right\rangle
+\left\vert 10\right\rangle -\left\vert 11\right\rangle \right)  .
\end{equation}
This is a typical quantum phase gate for each electron pairs. For example,
when the charge detectors P$_{1}$ and P$_{2}$ are both equal to 1, the output
qubits 1, 2, 3 and 4$^{^{\prime}}$ fall to
\begin{align}
\left\vert \Psi\right\rangle _{1234^{^{\prime}}}  &  =1/4\text{ }(\left\vert
0000\right\rangle +\left\vert 0001\right\rangle +\left\vert 0010\right\rangle
-\left\vert 0011\right\rangle \nonumber\\
&  +\left\vert 0100\right\rangle +\left\vert 0101\right\rangle -\left\vert
0110\right\rangle +\left\vert 0111\right\rangle \nonumber\\
&  +\left\vert 1000\right\rangle +\left\vert 1001\right\rangle +\left\vert
1010\right\rangle -\left\vert 1011\right\rangle \nonumber\\
&  -\left\vert 1100\right\rangle -\left\vert 1101\right\rangle +\left\vert
1110\right\rangle -\left\vert 1111\right\rangle ),
\end{align}
which can also be rewritten to be a standard form of the cluster state,%
\begin{align}
\left\vert \Psi\right\rangle _{1234^{^{\prime}}}  &  =1/4\text{ }\left(
\left\vert 0\right\rangle _{1}+\left\vert 1\right\rangle _{1}\sigma_{z}%
^{2}\right)  \otimes\left(  \left\vert 0\right\rangle _{2}+\left\vert
1\right\rangle _{2}\sigma_{z}^{3}\right) \nonumber\\
&  \otimes\left(  \left\vert 0\right\rangle _{3}+\left\vert 1\right\rangle
_{3}\sigma_{z}^{4}\right)  \otimes\left(  \left\vert 0\right\rangle
_{4}+\left\vert 1\right\rangle _{4}\right)  .
\end{align}

If P$_{1}$ and P$_{2}$ are not both equal to 1, a single-qubit operation is
needed after above operations to realize the required state in Eq. (4). For
example, in Table 1,\ where the first row means the case when $P_{1}$ gets the
value 0 or 1 and the first column means the case of $P_{2}$ = 0 or 1;
$\sigma_{x2^{\prime}}$ and $\sigma_{x3^{\prime}}$ are spin-flip operations on
the qubits output from the ports Nos. 2$^{\prime}$ and 3$^{\prime}$,
respectively, and $I$ is the identity operator. It is easy to check that above
idea can be extended to generation of N-electron cluster state by the design
shown in Fig. 2.\ For any P$_{j}$ = 0 we just need a spin-flip ($\sigma_{x}$)
operation on one of the electron spins in the control-out arm of the
corresponding encoders.

Now we start to construct analyzers for multipartite entangled states. In Fig.
3 we show how multipartite entangled states can be distinguished using blocks
of the encoder. In the full Bell-state analyzer\ shown in Fig. 3 (a), the
different outputs of $P_{j}$ (j = 1, 2) correspond to different Bell-states,
where $P_{1}=1\rightarrow\left\vert \Phi^{+}\right\rangle $ or $\left\vert
\Phi^{-}\right\rangle ;$ $P_{1}=0\rightarrow\left\vert \Psi^{+}\right\rangle $
or $\left\vert \Psi^{-}\right\rangle $, $P_{2}=1\rightarrow\left\vert \Phi
^{+}\right\rangle $ or $\left\vert \Psi^{+}\right\rangle ;$ $P_{2}%
=0\rightarrow\left\vert \Phi^{-}\right\rangle $ or $\left\vert \Psi
^{-}\right\rangle $, with $\left\vert \Phi^{\pm}\right\rangle $ and
$\left\vert \Psi^{\pm}\right\rangle $ being, respectively, the Bell-states
$\left\vert \Phi^{\pm}\right\rangle =\frac{1}{\sqrt{2}}\left(  \left\vert
00\right\rangle \pm\left\vert 11\right\rangle \right)  $ and $\left\vert
\Psi^{\pm}\right\rangle =\frac{1}{\sqrt{2}}\left(  \left\vert 01\right\rangle
\pm\left\vert 10\right\rangle \right)  .$ For example, if we have $P_{1}=1$
and $P_{2}=1$, we know the electrons passing our analyzer to be in $\left\vert
\Phi^{+}\right\rangle .$

In the case of a three-electron entangled state shown in Fig. 3 (b), after
qubits 1 and 2 go through the encoder $P_{1}$, \ we can distinguish \{$g_{1}$,
$g_{2}$\} from \{$g_{3}$, $g_{4}$\} when $P_{1}=1$ by a parity check on the
detected pairwise 1-2, where $g_{1}=\left\vert \psi\right\rangle _{i,ii}$,
$g_{2}=\left\vert \psi\right\rangle _{iii,iv}$, $g_{3}=\left\vert
\psi\right\rangle _{v,vi}$, $g_{4}=\left\vert \psi\right\rangle _{vii,viii}$,
by using following equations,
\begin{align}
\left\vert \psi\right\rangle _{i,ii}  &  =\frac{1}{\sqrt{2}}\left(  \left\vert
000\right\rangle _{123}\pm\left\vert 111\right\rangle _{123}\right)
\nonumber\\
&  =\left\vert \Phi^{+}\right\rangle _{12}\left\vert \pm\right\rangle
_{3}+\left\vert \Phi^{-}\right\rangle _{12}\left\vert \mp\right\rangle
_{3},\nonumber\\
\left\vert \psi\right\rangle _{iii,iv}  &  =\frac{1}{\sqrt{2}}\left(
\left\vert 110\right\rangle _{123}\pm\left\vert 001\right\rangle _{123}\right)
\nonumber\\
&  =\left\vert \Phi^{+}\right\rangle _{12}\left\vert \pm\right\rangle
_{3}-\left\vert \Phi^{-}\right\rangle _{12}\left\vert \mp\right\rangle
_{3},\nonumber\\
\left\vert \psi\right\rangle _{v,vi}  &  =\frac{1}{\sqrt{2}}\left(  \left\vert
010\right\rangle _{123}\pm\left\vert 101\right\rangle _{123}\right) \\
&  =\left\vert \Psi^{+}\right\rangle _{12}\left\vert \pm\right\rangle
_{3}+\left\vert \Psi^{-}\right\rangle _{12}\left\vert \mp\right\rangle
_{3},\nonumber\\
\left\vert \psi\right\rangle _{vii,viii}  &  =\frac{1}{\sqrt{2}}\left(
\left\vert 100\right\rangle _{123}\pm\left\vert 011\right\rangle _{123}\right)
\nonumber\\
&  =\left\vert \Psi^{+}\right\rangle _{12}\left\vert \pm\right\rangle
_{3}-\left\vert \Psi^{-}\right\rangle _{12}\left\vert \mp\right\rangle
_{3}.\nonumber
\end{align}
If P$_{1}$=0, with a $\sigma_{x}$ operation on one of the output electrons
from the encoder P$_{1},$ we can do the same job as in the case of P$_{1}$=1.
Similarly, after electrons go through the encoder P$_{2}$, we can distinguish
\{$g_{1}$, $g_{4}$\} from \{$g_{2}$, $g_{3}$\}. So up to now we have separated
the eight three-qubit entangled states into four groups, i.e., $g_{1}$,
$g_{2},g_{3}$, $g_{4}$. In order to distinguish the two states (i.e., the +
state from - state) in each groups, we have to perform measurements on each
group. For example, to distinguish the state $\frac{1}{\sqrt{2}}\left(
\left\vert 000\right\rangle +\left\vert 111\right\rangle \right)  $ from
$\frac{1}{\sqrt{2}}\left(  \left\vert 000\right\rangle -\left\vert
111\right\rangle \right)  ,$ if we get a click in the basis $\left\vert
+\right\rangle $ and P$_{3}$=1, or get a click in $\left\vert -\right\rangle $
and P$_{3}$=0, we have the state $\frac{1}{\sqrt{2}}\left(  \left\vert
000\right\rangle +\left\vert 111\right\rangle \right)  $. Otherwise we have
$\frac{1}{\sqrt{2}}\left(  \left\vert 000\right\rangle -\left\vert
111\right\rangle \right)  $. Although each measurement would destroy a qubit,
as the eight entangled states can be completely distinguished, the device in
Fig. 3 (b) would be useful for teleportation and superdense coding with
tripartite states \cite{19,20}.

The idea can be generalized to many-electron cases, which would be more
complicated but still deterministic. We show in Fig. 3 (c) an example for a
four-electron entanglement analyzer. By checking the parity of the two pairs
1-2 and 3-4, along with the readouts of P$_{1}$ and P$_{2},$ we can separate
the four-qubit entangled states into four sets of groups, as shown in Table 2,
by using following Eq. (6),
\begin{align}
\left\vert \psi\right\rangle _{i}  &  =\frac{1}{\sqrt{2}}\left(  \left\vert
0000\right\rangle _{1234}\pm\left\vert 1111\right\rangle _{1234}\right)
\nonumber\\
&  =\{%
\genfrac{}{}{0pt}{}{\frac{1}{\sqrt{2}}\left(  \left\vert \Phi^{+}\right\rangle
_{12}\left\vert \Phi^{+}\right\rangle _{34}+\left\vert \Phi^{-}\right\rangle
_{12}\left\vert \Phi^{-}\right\rangle _{34}\right)  \text{ for + state}%
}{\frac{1}{\sqrt{2}}\left(  \left\vert \Phi^{+}\right\rangle _{12}\left\vert
\Phi^{-}\right\rangle _{34}+\left\vert \Phi^{-}\right\rangle _{12}\left\vert
\Phi^{+}\right\rangle _{34}\right)  \text{ for - state}}%
,\nonumber\\
\left\vert \psi\right\rangle _{ii}  &  =\frac{1}{\sqrt{2}}\left(  \left\vert
0001\right\rangle \pm\left\vert 1110\right\rangle \right) \nonumber\\
&  =\{%
\genfrac{}{}{0pt}{}{\frac{1}{\sqrt{2}}\left(  \left\vert \Phi^{+}\right\rangle
_{12}\left\vert \Psi^{+}\right\rangle _{34}+\left\vert \Phi^{-}\right\rangle
_{12}\left\vert \Psi^{-}\right\rangle _{34}\right)  \text{ for + state}%
}{\frac{1}{\sqrt{2}}\left(  \left\vert \Phi^{+}\right\rangle _{12}\left\vert
\Psi^{-}\right\rangle _{34}+\left\vert \Phi^{-}\right\rangle _{12}\left\vert
\Psi^{+}\right\rangle _{34}\right)  \text{ for - state}}%
,\nonumber\\
\left\vert \psi\right\rangle _{iii}  &  =\frac{1}{\sqrt{2}}\left(  \left\vert
0010\right\rangle \pm\left\vert 1101\right\rangle \right) \nonumber\\
&  =\{%
\genfrac{}{}{0pt}{}{\frac{1}{\sqrt{2}}\left(  \left\vert \Phi^{+}\right\rangle
_{12}\left\vert \Psi^{+}\right\rangle _{34}-\left\vert \Phi^{-}\right\rangle
_{12}\left\vert \Psi^{-}\right\rangle _{34}\right)  \text{ for + state}%
}{\frac{1}{\sqrt{2}}\left(  \left\vert \Phi^{+}\right\rangle _{12}\left\vert
\Psi^{-}\right\rangle _{34}-\left\vert \Phi^{-}\right\rangle _{12}\left\vert
\Psi^{+}\right\rangle _{34}\right)  \text{ for - state}}%
,\nonumber\\
\left\vert \psi\right\rangle _{iv}  &  =\frac{1}{\sqrt{2}}\left(  \left\vert
0100\right\rangle \pm\left\vert 1011\right\rangle \right) \nonumber\\
&  =\{%
\genfrac{}{}{0pt}{}{\frac{1}{\sqrt{2}}\left(  \left\vert \Psi^{+}\right\rangle
_{12}\left\vert \Phi^{+}\right\rangle _{34}+\left\vert \Psi^{-}\right\rangle
_{12}\left\vert \Phi^{-}\right\rangle _{34}\right)  \text{ for + state}%
}{\frac{1}{\sqrt{2}}\left(  \left\vert \Psi^{+}\right\rangle _{12}\left\vert
\Phi^{-}\right\rangle _{34}+\left\vert \Psi^{-}\right\rangle _{12}\left\vert
\Phi^{+}\right\rangle _{34}\right)  \text{ for - state}}%
,\\
\left\vert \psi\right\rangle _{v}  &  =\frac{1}{\sqrt{2}}\left(  \left\vert
1000\right\rangle \pm\left\vert 0111\right\rangle \right) \nonumber\\
&  =\{%
\genfrac{}{}{0pt}{}{\frac{1}{\sqrt{2}}\left(  \left\vert \Psi^{+}\right\rangle
_{12}\left\vert \Phi^{+}\right\rangle _{34}-\left\vert \Psi^{-}\right\rangle
_{12}\left\vert \Phi^{-}\right\rangle _{34}\right)  \text{ for + state}%
}{\frac{1}{\sqrt{2}}\left(  \left\vert \Psi^{+}\right\rangle _{12}\left\vert
\Phi^{-}\right\rangle _{34}-\left\vert \Psi^{-}\right\rangle _{12}\left\vert
\Phi^{+}\right\rangle _{34}\right)  \text{ for - state}}%
,\nonumber\\
\left\vert \psi\right\rangle _{vi}  &  =\frac{1}{\sqrt{2}}\left(  \left\vert
0011\right\rangle \pm\left\vert 1100\right\rangle \right) \nonumber\\
&  =\{%
\genfrac{}{}{0pt}{}{\frac{1}{\sqrt{2}}\left(  \left\vert \Phi^{+}\right\rangle
_{12}\left\vert \Phi^{+}\right\rangle _{34}-\left\vert \Phi^{-}\right\rangle
_{12}\left\vert \Phi^{-}\right\rangle _{34}\right)  \text{ for + state}%
}{\frac{1}{\sqrt{2}}\left(  \left\vert \Phi^{+}\right\rangle _{12}\left\vert
\Phi^{-}\right\rangle _{34}-\left\vert \Phi^{-}\right\rangle _{12}\left\vert
\Phi^{+}\right\rangle _{34}\right)  \text{ for - state}}%
,\nonumber\\
\left\vert \psi\right\rangle _{vii}  &  =\frac{1}{\sqrt{2}}\left(  \left\vert
0101\right\rangle \pm\left\vert 1010\right\rangle \right) \nonumber\\
&  =\{%
\genfrac{}{}{0pt}{}{\frac{1}{\sqrt{2}}\left(  \left\vert \Psi^{+}\right\rangle
_{12}\left\vert \Psi^{+}\right\rangle _{34}+\left\vert \Psi^{-}\right\rangle
_{12}\left\vert \Psi^{-}\right\rangle _{34}\right)  \text{ for + state}%
}{\frac{1}{\sqrt{2}}\left(  \left\vert \Psi^{+}\right\rangle _{12}\left\vert
\Psi^{-}\right\rangle _{34}+\left\vert \Psi^{-}\right\rangle _{12}\left\vert
\Psi^{+}\right\rangle _{34}\right)  \text{ for - state}}%
,\nonumber\\
\left\vert \psi\right\rangle _{viii}  &  =\frac{1}{\sqrt{2}}\left(  \left\vert
1001\right\rangle \pm\left\vert 0110\right\rangle \right) \nonumber\\
&  =\{%
\genfrac{}{}{0pt}{}{\frac{1}{\sqrt{2}}\left(  \left\vert \Psi^{+}\right\rangle
_{12}\left\vert \Psi^{+}\right\rangle _{34}-\left\vert \Psi^{-}\right\rangle
_{12}\left\vert \Psi^{-}\right\rangle _{34}\right)  \text{ for + state}%
}{\frac{1}{\sqrt{2}}\left(  \left\vert \Psi^{+}\right\rangle _{12}\left\vert
\Psi^{-}\right\rangle _{34}-\left\vert \Psi^{-}\right\rangle _{12}\left\vert
\Psi^{+}\right\rangle _{34}\right)  \text{ for - state}}%
,\nonumber
\end{align}

where $\pm$ state means ($\left\vert \cdots\right\rangle $ $\pm$ $\left\vert
\cdots\right\rangle $)/$\sqrt{2}$, respectively. Another encoder P$_{3}$ would
further divide the four sets of groups into eight groups, i.e., $\left\vert
\psi\right\rangle _{j}$ (j=i, ii $\cdots$ viii). In order to completely
distinguish the sixteen four-qubit entangled states, we have to make two
measurements, as shown in Fig. 3 (c).

To achieve our proposal experimentally, we may encode the qubits into the spin
degrees of freedom of, for example, the conduction electrons in a quantum dot
system. A recent scheme \cite{14} for such conduction electrons is proposed to
convert spin parity into charge information by resonant tunneling between two
dots when the spins are antiparallel. Moreover, both the beam splitters and
the charge detectors required in our scheme have been experimentally achieved
\cite{21,22,23,24} by means of the point contacts in two-dimensional electron
gas. Since it only makes a parity check, the charge detector in our case can
be realized by using the point contact made of a quantum dot with a resonant
conductance. According to whether it is resonant or off-resonant, the detector
can distinguish occupation number one from occupation number 0 or two.
However, the great experimental challenge for charge detector is the time
resolved detection required for the ballistic electrons, which at present is
longer than our requirement by several orders \cite{24}. We expect that this
requirement could be met in the near future with the advance of the techniques
in this respect.

In conclusion, we have presented schemes for fermions to deterministically
generate cluster states and to deterministically distinguish multipartite
entanglement based on polarizing beam splitters and single-spin rotations in
combination with the charge detectors. The generation of the cluster states
would be useful for one-way quantum computing with fermions, and the
construction of analyzers for multipartite entangled state provides a
potential way for fermionic quantum information processing with multipartite
entangled states.

This work is supported in part by National Natural Science Foundation of China
under Grant Nos. 10474118 and 10274093, and partly by the National Fundamental
Research Program of China under Grant Nos. 2001CB309309 and 2005CB724502.

\end{document}